\documentstyle[11pt,epsfig,astron]{article}
\begin{document}
\begin{titlepage}
\title{The cosmological lens equation in a universe with non zerocosmological constant}
\author{{M. Demianski$^3$, R. de Ritis$^{1,2}$, A. A. Marino$^{2,4}$, E. Piedipalumbo$^1$}\\
{\em \small $^1$ Dipartimento di Scienze Fisiche, Universit\`{a} di Napoli,}\\
{\em \small Mostra d'Oltremare pad. 20-80125 Napoli, Italy;} \\
{\em\small$^2$Istituto Nazionale di Fisica Nucleare, Sezione di Napoli,}\\
{\em \small Complesso Universitario di Monte S. Angelo, Via Cinzia,
Edificio G 80126 Napoli, Italy;}\\ {\em \small $^3$Institute for
Theoretical Physics, University of Warsaw, Warsaw, Poland} \\
{\em\small$^4$Osservatorio Astronomico di Capodimonte,}\\{\em \small Via
Moiariello, 16-80131 Napoli, Italy}}.
\date{}
\maketitle
\begin{abstract}
We derive and solve exactly the Dyer-Roeder equation in a Friedman
Robertson Walker cosmological model with non zero cosmological constant. To
take into account non homogeneous distribution of matter we use the
phenomenological clumpiness parameter $\tilde{\alpha}$. We propose also a
general form of an approximate solution which is simple enough to be useful
in practical applications and sufficiently accurate in the interesting
range of redshifts.
\end{abstract}
\vspace{20.mm}
e-mail
addresses:\\mde@fuw.edu.pl\\deritis@na.infn.it\\marino@na.astro.it\\ester@na.infn.it\\
\vfill
\end{titlepage}
\section{Introduction}
Recent determination of distances to high redshift type Ia supernovae
revealed that the universe is now expanding faster  than expected. This
accelerated expansion is linked with a non zero cosmological constant.
Since gravitational lensing of cosmological objects involves also high
redshift galaxies it is interesting to extend the cosmological lens
equation to cosmological models with non zero cosmological
constant.\\Furthermore the mass distribution in the universe at the
relevant range of redshifts is clearly non homogeneous. In our
considerations we will take into account the fact that at the scale up to
several hundred Mpc matter distribution is non homogeneous. Moreover, with
the advent of the Next Generation Space Telescope (NGT), which will allow
imaging of objects at $z\geq10$ (\cite{Bark}),  it will be possible to
check better and better the influence of  non homogeneities
.\\Using the general equations describing propagation of light in a given
spacetime we derive an equation for the angular diameter distance and we
extend the notion of angular diameter distance to two objects at arbitrary
redshifts. The general equation for the angular diameter distance, the so
called Dyer-Roeder equation, can be exactly solved also in the case when
the cosmological constant is different from zero. However, the exact
solution is very complicated  and therefore not useful for practical
applications. We propose a simple analytic form of an approximate solution.
The approximate form of the angular diameter distance depends on four
arbitrary parameters. We  fix the values of these parameters by fitting
this form to the exact solution.\\The paper is organized as follows: In
Section $2$ we present the derivation of the angular diameter distance in
an arbitrary Friedman- Robertson- Walker cosmological model. To take into
account the non homogeneous distribution of matter, following the standard
practice, we introduce a constant phenomenological parameter
$\tilde{\alpha}$ and rewrite the final equation in a few different forms.
In Section $3$ we present exact analytic solutions for the angular diameter
distance, first in a flat universe and then in the general case of
arbitrary spatial curvature. Section $4$ is devoted to the approximate
solution and to  discussions of the fitting procedure used to determine the
values of arbitrary parameters. Discussion of our results is presented in
Section $5$.
\section{General considerations}

Let us consider a beam of light emanating from a source S.
The light rays propagate along a surface $\Sigma$ which is determined
by  the eikonal equation
\begin{equation}
g^{\alpha \beta}\Sigma,_{\alpha}\Sigma,_{\beta}=0
\label{eq: eiko}
\end{equation}
A light ray is identified with a null geodesic on $\Sigma$ with the
tangent vector $k_{\alpha}=-\Sigma,_{\alpha}$. The light rays in the beam
can be  described by $x^{\alpha}=x^{\alpha}(v, y^{a})$ where $v$ is an
affine parameter and $y^{a}$ (a=1, 2, 3) are three  parameters  specifying
different rays. The tangent vector field to the light  ray congruence,
$k^{\alpha}={dx^{\alpha}\over dv}=-\Sigma,_{\alpha}$, determines two
optical scalars, the expansion $\theta$ and the shear $\sigma$:
\begin{equation}
\theta={1\over 2}{k^{\alpha}}_{;\alpha}\; \ \ \ \
\sigma=k_{\alpha;\beta}{\bar m}^{\alpha}{\bar m}^{\beta},
\label{eq: opscal1}
\end{equation}
where ${\bar m}^{\alpha}={1\over \sqrt{2}}(\xi^{\alpha}-i\eta^{\alpha})$
is a complex vector spanning the spacelike 2-space (the screen space)
orthogonal to $k^{\alpha}$ $(k^{\alpha}{\bar m}_{\alpha}=0)$ (actually the
vorticity connected with the light beam is zero in all our
considerations, therefore in our case these two scalars completely
characterize the congruence). These two
optical scalars describe the relative rate of change of an infinitesimal
area A of the cross section of the beam of light rays and its distortion.
In particular
\begin{equation}
\theta={1\over 2}{k^{\alpha}}_{;\alpha}={1\over 2} {d\ln{A}\over
dv}.
\label{eq: opscal2}
\end{equation}
These two  optical scalars satisfy the Sachs (\cite{Sachs}) propagation
equations
\begin{equation}
\dot{\theta} +\theta^2 + |\sigma |^2 = -
\frac{1}{2} R_{\alpha \beta} k^{\alpha} k^{\beta},
\label{eq: raysachs1}
\end{equation}
and
\begin{equation}
\dot{\sigma} + 2\theta \sigma = - \frac{1}{2} C_{\alpha \beta \gamma \delta}
{\bar m}^{\alpha} k^{\beta}{\bar m}^{\gamma} k^{\delta},
\label{eq: raysachs2}
\end{equation}
where the dot denotes the derivative with respect to $v$,
$R_{\alpha\beta}$ is the Ricci tensor, and $C_{\alpha \beta \gamma\delta}$
is the Weyl tensor.  Equations (\ref{eq: raysachs1}) and (\ref{eq:
raysachs2}) follow from the Ricci identity. We will use these equations to
study propagation of light in the Friedman-Robertson-Walker (FRW)
spacetime. The FRW spacetime is conformally flat, so in such spacetimes
$C_{\alpha\beta\gamma\delta}=0$. From equation (\ref{eq: raysachs2}) it
follows that if initially the shear of the null ray congruence is
equal to
zero than it is always zero. Therefore assuming that the light beam
emanating from the source S has vanishing shear we can disregard the shear
parameter altogether. Using (\ref{eq: opscal2}) we can rewrite equation
(\ref{eq: raysachs1}) in the form
\begin{equation}
\ddot{\sqrt{A}}+{1\over 2}R_{\alpha \beta}k^{\alpha}k^{\beta}\sqrt{A}=0.
\label{eq: raysachs3}
\end{equation}

An observer moving with the 4-velocity vector $u^{\alpha}$ will be
associated  with  the light ray  circular frequency
$\omega=cu^{\alpha}k_{\alpha}$. Different observers will assign different
frequencies to the same light ray. The shift of frequencies as measured by
an observer comoving with the sources and an arbitrary observer is related
to the redshift by
\begin{equation}
1+z={\omega \over \omega_{o}}={c \over
\omega_{o}}u^{\alpha}k_{\alpha},
\label{eq: redshift1}
\end{equation}
where $\omega_{o}$ is the frequency measured by the distant
observer. Differentiating this equation with respect to the affine
parameter $\it{v}$ we obtain
\begin{equation}
{dz\over dv}={c\over \omega_{o}}k^{\alpha}k^{\beta}u_{\alpha;\beta}.
\label{eq: dzdv}
\end{equation}

Since the angular diameter distance D is proportional to $\sqrt{A}$ we can
rewrite equation (\ref{eq: raysachs3}) using D instead of $\sqrt{A}$
and at the same time we
replace  the affine parameter $\it{v}$ by the redshift $z$, we obtain
\begin{equation}
\left ( \frac{dz}{dv} \right )^2 \frac{d^2 D}{dz^2} + \left (
\frac{d^2z}{dv^2} \right ) \frac{dD}{dz} +  \frac{4\pi G}{c^{4}} T_{\alpha
\beta} k^{\alpha} k^{\beta} D = 0,
\label{eq: angdiam2}
\end{equation}
where we used the Einstein equations to replace the Ricci tensor by
the energy-momentum tensor.

To relate a solution of (\ref{eq: angdiam2}) with the distance it has
to satisfy the following initial conditions:
\begin{eqnarray}
&&D(z)\arrowvert_{z = 0} = 0, \nonumber\\ && \\ &&\frac{dD(z)}{dz}
\arrowvert_{z = 0} = \frac{c}{H_0}.\nonumber
\label{eq: initialcond}
\end{eqnarray}

To be able to use solutions of the Eq. (\ref{eq: angdiam2}) to describe
gravitational lenses we have to introduce the distance between the source
and the lens $D(z_{l}, z_{s})$, where $z_{l}$ and $z_{s}$ denote
correspondingly the redshift of the lens and the source. Let
$D(z_{1},z_{2})$ denote the angular diameter distance between a fictitious
observer at $z_{1}$ and a source at $z_{2}$,\, of course $D(0,z)=D(z)$.
Suppose that we  know the general solution  of equation (\ref{eq:
angdiam2}) for $D(z)$ which satisfies the initial conditions (\ref{eq:
initialcond}), then the function $D(z_{1},z)$ defined by
\begin{equation} D(z_1, z) = \frac{c}{H_0}(1+z_1) D(z_1) D(z)
\left| \int_{z_1}^{z}{\frac{dz'}{D^2(z') g(z')}}\right| \ ,
\label{eq: dzunoz}
\end{equation}
such that

\begin{eqnarray}
&&D(z_1, z) \arrowvert_{z= z_1} = 0, \nonumber \\ &&\nonumber\\
&&\frac{d}{dz} D(z_1, z) \arrowvert_{z = z_1} = {\rm sign}(z - z_1)
\frac{1+z_1}{g(z_1)}, \nonumber
\label{eq : rz1z2}
\end{eqnarray}

satisfies equation (\ref{eq: angdiam2}), if the function $g(z)$ is a solution of
\begin{eqnarray}
\frac{d}{dz} \ln{g(z)} =
\frac{\frac{d^2z}{dv^2}}{\left ( \frac{dz}{dv} \right )^2},
\label{eq: gzeq}
\end{eqnarray}
so
\begin{equation} g(z) = g_0
\exp{\int{\frac{\frac{d^2z'}{dv^2}}{\left ( \frac{dz'}{dv} \right
)^2}dz'}}. \label{eq: gz}
\end{equation}
$g_{0}$  is an arbitrary constant of integration, which, without
restricting generality, we assume to be one. To obtain the equation
(\ref{eq: gzeq}) we inserted (\ref{eq: dzunoz}) into the equation (\ref{eq:
angdiam2}) and demanded that it is satisfied. Let us note that the
Etherington reciprocity relation (\cite{Ether33})
\begin{equation}
\frac{D(z_1, z_2)}{1 + z_1} = \frac{D(z_2, z_1)}{1
+ z_2}, \label{eq: ether}
\end{equation}
follows directly from (\ref{eq: dzunoz}).

We are interested in applying  equation (\ref{eq: angdiam2}) to
find the angular diameter distance to objects at high redshifts. Therefore let
us consider the standard FRW spacetime described by the line element
\begin{equation}
ds^2 = dt^2 - R^2(t) \left [
\frac{dr^2}{1-k r^2} + r^2(d\theta^2 + \sin^2{\theta} d\varphi^2)
\right ],
\label{eq: linefrw}
\end{equation}
where k (-1, 0, +1) is the curvature constant, and $R(t)$ is the scale
factor. If the matter content of the universe can be approximated by dust
then, as shown by Friedman, the Einstein field equations for the metric
(\ref{eq: linefrw}) assume the form
\begin{eqnarray}
&&\frac{{\dot{R}}^{2}}{R^{2}}= H^2 = \frac{8\pi G}{3} \varrho_{m} +
\frac{\Lambda c^2}{3} -
\frac{kc^{2}}{R^2}, \nonumber\\
\label{eq:einst}
&&\\ &&\frac{\ddot{R}}{R} = - \frac{4\pi G}{3}
\varrho_{m} + \frac{\Lambda c^{2}}{3}\nonumber
\end{eqnarray}
where here the dot denotes derivative with respect to $t$, $H=
\displaystyle{\frac{\dot{R}}{R}}$, $\varrho_{m}$
 is the matter density,
and $\Lambda$ is the cosmological constant.

In the FRW spacetime the redshift $z$  is connected with the scale
factor $R(t)$ by $1+z=\displaystyle{{R_{0}\over R(t)}}$. Differentiating this
relation with respect to time we obtain
\begin{eqnarray}
{dz\over dt}=-(1+z)H(z).
\nonumber \\
\nonumber
\end{eqnarray}

Let us now return to the description of propagation of a light beam.
Assuming that the observer is a standard FRW observer, e.g. it is comoving
with matter, from the equation (\ref{eq: dzdv}), we obtain
\begin{equation}
{c^{2}\over \omega_{0}}{dz\over dv}=(1+z)^{2}H(z).
\label{eq: dem}
\end{equation}

Introducing a new dimensionless affine parameter
$w= {H_{0}\omega_{0}\over c^{2}}v$  we
transform  equation  (\ref{eq: dem}) into
\begin{equation}
H_{0}{dz\over dw}=(1+z)^{2}H(z)
\label{eq:hubble1}
\end{equation}

The  Friedman equation for $H^2$, Eq. (\ref{eq:einst}), can be rewritten as
\begin{equation}
H^{2}(z)=H_{0}^{2}\left(\Omega_{m}(1+z)^{3}+\Omega_{k}(1+z)^{2}+
\Omega_{\Lambda}\right)
\label{eq:hubble2}
\end{equation}
where $H_{0}$ is the present value of the Hubble constant, and
\begin{displaymath}
\Omega_{m}={{8\pi G\varrho_{0}}\over {3H_{0}^{2}}}, \ \ \ \
\Omega_{k}=-{kc^{2}\over R_{0}^{2}H_{0}^{2}}, \ \ \
\Omega_{\Lambda}={{\Lambda c^{2}}\over 3H_{0}^{2}},
\end{displaymath}
are the present density parameters of matter, curvature and cosmological
constant respectively, and, in the case $p=0$ (no radiation), we have
\begin{equation}
\Omega_{m}+\Omega_{k}+\Omega_{\Lambda}=1.
\end{equation}
Substituting (\ref{eq:hubble2}) into (\ref{eq:hubble1}) we get
\begin{equation}
{dz\over dw}=(1+z)^{2}\sqrt{\Omega_{m}(1+z)^{3}+\Omega_{k}(1+z)^{2}+
\Omega_{\Lambda}}.
\end{equation}

To apply the angular diameter distance to the realistic distribution of
galaxies it is necessary to take into account local non homogeneities.
Unfortunately so far an acceptable averaging procedure for smoothing out
local non homogeneities has not been developed (\cite{Kraz}). Therefore
following previous discussions we introduce the phenomenological parameter
$\tilde{\alpha}$ which describes the influence of local non homogeneities
on propagation of light (\cite{DR72}), (\cite{PRT99}). With this
alteration equation (\ref{eq: angdiam2}), in the FRW-dust case, can be
rewritten  in the form
\begin{equation}
\left ( \frac{dz}{dw} \right )^2 \frac{d^2D}{dz^2} +
\left ( \frac{d^2z}{dw^2} \right ) \frac{dD}{dz} +
\frac{3}{2} \tilde{\alpha} \Omega_{m} (1+z)^5 D = 0 .
\label{eq:angdiamalpha}
\end{equation}
It is customary to measure cosmological distances in units of ${c\over
H_{0}}$,
introducing the dimensionless angular diameter distance (the
Dyer-Roeder distance) $r={DH_{0}\over c}$,  and using
(\ref{eq:angdiamalpha}) we finally obtain
\begin{eqnarray}
&&(1+z)\left[
\Omega_{m}(1+z)^{3}+\Omega_{k}(1+z)^{2}+\Omega_{\Lambda}\right]  {d^{2}r\over
dz^{2}}  +\nonumber \label{eq: stdr2}\\ &&  \\ &&\left({7\over
2}\Omega_{m}(1+z)^{3}+3\Omega_{k}(1+z)^{2}+2\Omega_{\Lambda}\right)
{dr\over dz} +{3\over 2}{\tilde \alpha}\Omega_{m}(1+z)^{2}r  = 0.
\nonumber
\end{eqnarray}
with the initial conditions:
\begin{eqnarray}
&&r(z)\arrowvert_{z = 0} = 0, \nonumber\\ && \\ &&\frac{dr(z)}{dz}
\arrowvert_{z = 0} = 1.\nonumber
\label{eq: initialcond2}
\end{eqnarray}

This equation can be cast into a different form by using $a={R\over R_{0}}$
as a parameter instead of $z$, we obtain
\begin{equation}
a^{2}(\Omega_{m}+\Omega_{k}a+\Omega_{\Lambda}a^{3}){d^{2}r\over da^{2}}-
a({3\over 2}\Omega_{m}+\Omega_{k}a){dr\over da}+ {3\over
2}{\tilde{\alpha}}\Omega_{m}r=0,
\label{eq: scaledr}
\end{equation}
or using the cosmic time $t$ as a parameter, the equation
(\ref{eq:angdiamalpha}) assumes the form
\begin{equation}
{d^{2}r\over dt^{2}}-H(t){dr\over dt}+
4\pi G{\tilde \alpha} \varrho_{m}(t)r=0.
\label{eq: tempodr}
\end{equation}

Equation (\ref{eq: tempodr}) needs some comments: this equation was first
introduced by Dashevski \& Zeldovich (\cite{DZ65}) (see also Dashevski
\& Slysh, \cite{DS66}). More recently Kayser et al.
(\cite{KHS97}) have used it to derive an equation similar to our (\ref{eq:
stdr2}).  In Eq.(\ref{eq: tempodr}) the clumpiness parameter
$\tilde{\alpha}$ is usually considered as a constant. However in the
papers by Dashevski \& Zeldovich and  by Kayser et al. $\tilde{\alpha}$ is
allowed to vary with time but only the case $\tilde{\alpha}=const.$ is
really considered. For a discussion of the case in which $\tilde{\alpha}$
depends on $z$ see, for example, the paper by Linder (\cite{Lin88}).

As an  example of our procedure let us consider the case when
$\Omega_{\Lambda} =0$ and $\Omega_{k}=0$, that is a flat Universe so the
universe is flat. In this case it is easy to see that
\begin{equation}
r(z)={{(1+z)^{\beta}-(1+z)^{-\beta}}\over {2\beta (1+z)^{5/4}}},
\label{eq: SEFsol}
\end{equation}
where $\beta={1\over 4}\sqrt{25-24\tilde{\alpha}}$, is the general
solution of equation (\ref{eq: stdr2}) with appropriate initial
conditions. This solution can be found in SEF (\cite{SEF}), see their
equation (4.56). The $g$ function can be easily deduced in the flat
$\Lambda=0$ Friedman-Robertson-Walker cosmological model and we obtain
\begin{eqnarray}
g(z) = (1+z)^2 \sqrt{\Omega_{M_0}z + 1}.\nonumber
\end{eqnarray}
Substituting the expression for $g(z)$ we get the familiar solution
$D(z_{l}, z_{s})$ found in SEF.
\section{Exact solutions }

In  recent papers Kantowski (\cite{Kant98}),(\cite{Kant20})  has found the
general solution of the Dyer-Roeder equation written in the form (\ref{eq:
stdr2}). In what follow we use the DR equation in the form (\ref{eq:
scaledr}) and obtain the general solution using more direct approach.

\subsection{The case $\Omega_k =0$}

When $\Omega_{k}=0$, it is possible to divide the
Eq.(\ref{eq: scaledr})
by $\Omega_{m}$ and it becomes\,:
\begin{equation}
a^2 (1 + \mu a^3)\frac{d^{2}\bar{r}}{da^{2}} - \frac{3}{2}
a\frac{d\bar{r}}{da} +
\frac{3}{2}
\tilde{\alpha} \bar{r} = 0 \ ,
\label{eq: scaledr2}
\end{equation}
where $\mu=\displaystyle{\Omega_{\Lambda}\over \Omega_{m}}$.

To solve Eq.(\ref{eq: scaledr2}) we use the following  strategy: First we
solve this equation in the case when $\mu=0$ and we look for solutions in
the power form $a^{s}$. Inserting this form into (\ref{eq: scaledr2}), we
obtain:
\begin{equation}
2s^2 - 5s + 3 \tilde{\alpha} = 0,
\label{eq: rs1}
\end{equation}
which has the solutions :
\begin{eqnarray}
&&s_{\pm}  = \frac{5}{4} {\pm} \frac{1}{4}\sqrt{25 - 24 \tilde{\alpha}} =
\frac{5}{4} {\pm} {\beta}, \nonumber\\ & & \\
&&\beta  = {\frac{1}{4}}\sqrt{25 - 24 \tilde{\alpha}}. \nonumber
\label{eq: s+-}
\end{eqnarray}
Writing the general solution is terms of ${\it z}$ instead of ${\it a}$
and imposing the initial conditions (\ref{eq: initialcond}) we recover the
solution (\ref{eq: SEFsol}).

In the general case when $\mu\not=0$ we look for solutions in the form
$\bar{r}_{\Lambda}=a^{s}f(x)$ where $x=a^{3}$. Inserting this form into
(\ref{eq: scaledr2}) after some rearrangements we obtain
\begin{equation}
3x (1 + \mu x) \frac{d^2 f}{dx^2} + \left ( 2(s+1)(1 + \mu x)-\frac{3}{2} \right ) \frac{df}{dx} +
\frac{\mu}{2}(s- \tilde{\alpha}) f = 0.
\label{eq: gauss}
\end{equation}
This equation can be reduced to  the standard hypergeometric equation.
The general solution of  Eq. (\ref{eq: scaledr2}) can be written in
the form\,:
\begin{eqnarray}
\bar{r}_{\Lambda} = A_1\frac{(1+z)^{-\beta}}{(1+z)^{5/4}}f_{s_{+}}
\left( \left (\frac{1}{1+z} \right )^3 \right ) &+& \nonumber\\ &&\\ A_2
\frac{(1+z)^{\beta}}{(1+z)^{5/4}}f_{s_{-}} \left( \left (
\frac{1}{1+z} \right )^3 \right ),\nonumber
\label{eq: venti9} \end{eqnarray}
where $A_1, A_2$ are arbitrary constants and we denoted the solutions by
$\bar{r}_{\Lambda}$ to stress that it is the solution of the DR equation
in a spacetime with $\Lambda\not=0$. Here $f_{s_{-}}$ and $f_{s_{+}}$ are
solutions of Eq.(\ref{eq: scaledr2}) with $s=s_{+}$ and $s=s_{-}$
correspondingly. The constants $A_{1}$ and $A_{2}$ are determined from the
initial conditions (\ref{eq: initialcond}), or explicitly:

\begin{eqnarray}
{\bar{r}}_{\Lambda}(z)\arrowvert_{z = 0} &=& A_1 f_{s_{+}}(1) + A_2
f_{s_{-}}(1)= 0,\nonumber \\ &&  \\
\frac{d{\bar{r}}_{\Lambda}}{dz}\arrowvert_{z = 0} &=& - s_{+} A_1 f_{s_{+}}(1) + A_1 \frac{d}{dz} f_{s_{+}}\arrowvert_{z = 0}
- s_{-} A_2 f_{s_{-}}(1) +\nonumber\\&&A_2 \frac{d}{dz}
f_{s_{-}}\arrowvert_{z = 0}= 1.\nonumber
\end{eqnarray}
To  find the function ${\bar{r}}_{\Lambda}(z_1,z_2)$ we have to solve the
equation (\ref{eq: gzeq}), which in this case is:
\begin{equation}
\frac{{{g}_{\Lambda}}'}{{g}_{\Lambda}}= - \frac{3}{2a(1+ \mu
a^3)}
\end{equation}
from which we get:
\begin{equation}
{g}_{\Lambda} = \left [(1+z)^3 + \mu \right ]^{1/2}.
\end{equation}
Then the general two points ($z>z_1$) solution of Eq.(\ref{eq: scaledr2})
is\,:
\begin{eqnarray}
{\bar{r}}_{\Lambda}(z_1, z)
&=&{\bar{r}}_{\Lambda}(z_1)(1+z_1){\bar{r}}_{\Lambda}(z){\times}\nonumber \\&&\\
&&\int_{z_1}^{z}{\frac{\sqrt{\Omega_m}dz'} {{\bar{r}}_{\Lambda}^2 (z')
(1+z')^2 [ (1+z')^3 \Omega_m +\Omega_{\Lambda}
]^{1/2}}},\nonumber\label{eq: rlambdaz1z2}
\end{eqnarray}
which satisfies the initial conditions:
\begin{eqnarray}
&&{\bar{r}}_{\Lambda}(z_1, z_1) = 0, \nonumber\\ &&\\
&&\frac{d{\bar{r}}_{\Lambda}}{dz}(z_1,z) \arrowvert_{z = z_1} =
\frac{{\rm sign} (z - z_1)\sqrt{\Omega_m}}{(1+z_1) \sqrt{(1+z)^3 \Omega_m +
\Omega_{\Lambda}}}.\nonumber
\end{eqnarray}
(when $\Omega_{\Lambda} = 0$, we get exactly the SEF solution (4.53) for
$\Omega_m=1$). It is worth to stress that using  properties of
the hypergeometric functions $(f_{s_{+}},f_{s_{-}})$ it turns out  that,
\begin{equation}
{\bar{r}} < {\bar{r}}_{\Lambda},
\label{eq: relat}
\end{equation}
which is also shown in the following figure.

\begin{figure}[ht]
\centering
\epsfig{figure=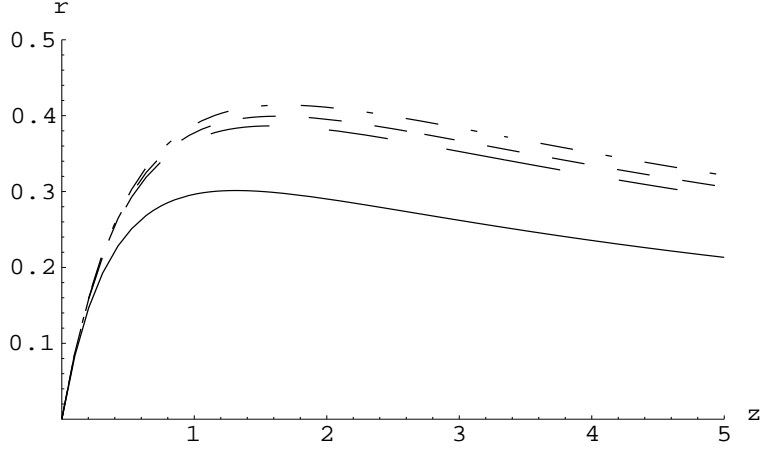,height=6cm,width=0.8\textwidth,clip=}
\caption{The figure shows  $r(z)$
for fixed $\tilde{\alpha}=0.9$  and for $\Omega_{\Lambda} = 0$,
$\Omega_{\Lambda} = 0.6$, $\Omega_{\Lambda }= 0.65$ and
$\Omega_{\Lambda}=0.7$. These figures have to be read from the bottom for
increasing values of $\Omega_{\Lambda}$. We see that
$\bar{r}<\bar{r_\Lambda}$ as clarified by Eq. (\ref{eq: relat})}
\end{figure}

\begin{figure}[ht]
\centering
\epsfig{figure=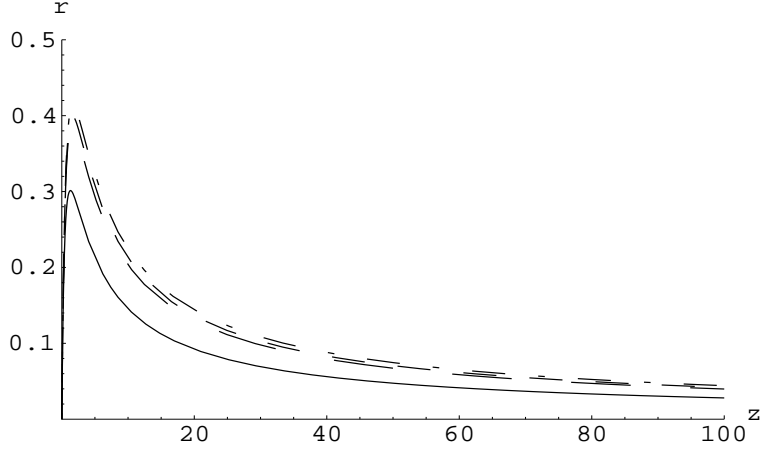,height=6cm,width=0.8\textwidth,clip=}
\caption{Same as Fig1, the z-range is now bigger ($0$, $100$)}
\end{figure}

\subsection{The case $\Omega_k \ne 0$}

When $\Omega_k \ne 0$ the Eq.(\ref{eq: scaledr}) can be rewritten in the
following form\,:

\begin{eqnarray}
&&\frac{d^2 r}{da^2} - \frac{\delta a +\frac{3}{2}}{a(a - a_1) (a -
a_2)(a -\bar{a}_2)} \frac{dr}{da}+\nonumber\\ &&\\ &&+ \frac{3}{2}
\tilde{\alpha}\frac{1}{a^2 (a - a_1) (a - a_2) (a -
\bar{a}_2)}r = 0,\nonumber \label{eq: drknz}
\end{eqnarray}
where $a_1, a_2, a_2$ are the roots of the equation $\Omega_m +\Omega_k a+
\Omega_{\Lambda}  a^3 = 0$, and $\delta=\frac{\Omega_{k}}{\Omega_{m}}$
(we are using the symbol $r$ for the Dyer-Roeder distance in this more
general case).
 These roots are\,:
\begin{displaymath} a_1 =
- \frac{\left ( \frac{2}{3} \right )^{1/3}\delta}{\sqrt{\mu} (-9
\sqrt{\mu} + \sqrt{3} \sqrt{4 \delta^3 + 27\mu})^{1/3}} + \frac{(-9 \sqrt{\mu}
+ \sqrt{3} \sqrt{4 \delta^3 + 27\mu})^{1/3}}{2^{1/3} 3^{2/3} \sqrt{\mu}},
\;
\end{displaymath} \begin{displaymath}
a_2 = \frac{(1 + i \sqrt{3})\delta}{2^{2/3} 3^{1/3} \sqrt{\mu} (-9 \sqrt{\mu}
+ \sqrt{3} \sqrt{\delta^3 + 27 \mu})^{1/3}} - \end{displaymath}
\begin{displaymath}
\ \ \\- \frac{(1 - i \sqrt{3})(-9 \sqrt{\mu} + \sqrt{3} \sqrt{4 \delta^3 +
 27\mu})^{1/3}}{2 2^{1/3} 3^{2/3} \sqrt{\mu}}, \;
\end{displaymath}
\begin{displaymath}
\bar{a}_2 = \frac{(1 - i \sqrt{3})\delta}{2^{2/3} 3^{1/3} \sqrt{\mu} (-9
\sqrt{\mu} + \sqrt{3} \sqrt{4 \delta^3 + 27 \mu})^{1/3}} -
\end{displaymath}
\begin{displaymath}
\ \ \\ - \frac{(1 + i \sqrt{3})(-9 \sqrt{\mu} + \sqrt{3} \sqrt{4\delta^3 + 27\mu})^{1/3}}{2 2^{1/3} 3^{2/3}
 \sqrt{\mu}} \ .
\end{displaymath}
where $\mu=\Omega_{\Lambda}/\Omega_n$ as before. ($a_2$ and $a_3$ are
complex conjugate and  $a_1 < 0$, then Eq. (\ref{eq: drknz}) has non
singular coefficients for real $a>0$). Equation  (\ref{eq: drknz}) is of
Fuchsian type (\cite{Ince},\cite{TRIC61})  with four regular singular
points plus a regular singular point at infinity. This equation can be put
in the following form
\begin{eqnarray}
\frac{d^2r}{da^2}-\left(\frac{\tilde{A}}{a}+\frac{\tilde{B}}{a-a_1}+
\frac{\tilde{C}}{a-a_2}+ \frac{\tilde{D}}{a-a_3}\right)
\frac{dr}{da}&+&\nonumber\\ &&\\
\frac{1}{a}\left(\frac{A}{a}+\frac{B}{a-a_1}+\frac{C}{a-a_2}+
\frac{D}{a-a_3}\right)r=0,\nonumber
\label{eq: reqstand}
\end{eqnarray}
where the coefficient A, B, C, D,  $\tilde{A}$,  $\tilde{B}$,  $\tilde{C}$,
$\tilde{D}$ are easily found in terms of the cosmological parameters
$\Omega_m$, $\Omega_{\Lambda}$, $\Omega_k$ and $\tilde{\alpha}$, and are
given in Appendix A.

The solutions around  each of the $4+1$ singularity points ( here "$1$"
denotes the singularity point  relative to the infinity) are grouped
together using the so called {\it{Riemann-P symbol.}}

$$\mathcal{P}\left(
\begin{array}{cccccc}
 0 & a_1 & a_2 & a_3 &

\infty &  \  \\
 \delta_{11} & \delta_{21} &\delta_{31} & \delta_{41} &

\delta_{1}& a  \\
\delta_{12} &\delta_{22} & \delta_{32} &

\delta_{42}&\delta_{2}& \
\end{array} \right),
$$
 where the $\delta_{ij}$ and  $\delta_i $ are given by
\begin{eqnarray}
&&\delta_{11}=\frac{5}{2}-\sqrt{\frac{25}{4} -\frac{25\tilde{\alpha}}{4}}
\,\, (=2s_{-}),
\nonumber\\
& &  \nonumber \\ &&\delta_{12}=\frac{5}{2}+\sqrt{\frac{25}{4}-
\frac{25\tilde{\alpha}}{4}}\, \,(=2s_{+}), \nonumber\\
& &  \nonumber \\ &&\delta_{21}=0, \nonumber\\ & &  \nonumber \\
&&\delta_{22}=1+\frac{\frac{3}{2}\Omega_m+\Omega_k
a_2}{a_1(a_2-a_1)(a_2-a_3)},
\nonumber\\
& &  \nonumber \\ &&\delta_{31}=0,\nonumber\\ & &  \nonumber \\
&&\delta_{32}=1+\frac{\frac{3}{2}\Omega_m+\Omega_k
a_1}{a_2(a_2-a_1)(a_2-a_3)},
\nonumber\\
& &  \nonumber \\ &&\delta_{41}=0,\nonumber\\
&&\delta_{42}=1+\frac{\frac{3}{2}\Omega_m+\Omega_k
a_3}{a_3(a_3-a_1)(a_3-a_2)},
\nonumber\\
& &  \nonumber \\ &&\delta_1=0,\nonumber\\& &  \nonumber \\
&&\delta_2=1,\nonumber
\end{eqnarray}
and correspond, respectively, to the solutions of the indicial equation
relative to the finite and infinite regular singular points of the
equation. We do not discuss further properties of these solutions because
unfortunately they cannot be given in an explicit analytical form. In the
next section we will give an approximate analytical expression for the
solution of equation (\ref{eq: scaledr}).

\section{The approximate lens equation}

The exact solutions presented in the previous section are complicated and difficult to use in practical applications. Therefore we want to find an analytical approximate expression for
$r(z,\Omega_m,\Omega_{\Lambda},\Omega_k,\tilde{\alpha})$ and for the
function $\chi$ appearing in the cosmological lens equation. Following SEF let us
 briefly recall the basic equations used in the derivation of the cosmological lens  equation and
introduce appropriate notation. The time delay between different light rays
reaching the observer is:
\begin{equation}
c \Delta t = (1+z_d) \left\{
\frac{D_d D_s}{D_{ds}}(\vec{\theta} - \vec{\beta}) - \psi(\vec{\xi})\right \}
+ const,
\label{eq: timedelay}
\end{equation}
where $D_d$, $D_s$ and $D_{ds}$ are correspondingly the angular diameter
distances to deflector, source and the angular diameter distance between
deflector and source.

The first term in the bracket represents the geometrical time delay and the
second one is connected with the non homogeneous distribution of matter. To
describe this effect we use a perturbed metric in the form \,:
\begin{displaymath}
ds^2 = a^2(\eta)
\left \{\left ( 1 + \frac{2U}{c^2} \right ) d\eta^2- \left ( 1 -\frac{2U}{c^2} \right ) d\sigma^2
\right \},
\end{displaymath}
where $\eta$ is the conformal time, $U$ is the gravitational potential of the deflector,  and $d\sigma^2$ is the spatial line element.
The first term in Eq.(\ref{eq: timedelay}) can be rewritten as\,:
\begin{equation}
(1+z_d) \frac{D_d D_s}{D_{ds}} = \frac{c}{H_0} [ \chi(z_d) - \chi(z_s) ]^{-1} \ ,
\label{eq:firstpiece}
\end{equation}
where the function $\chi$ is given by\,:
\begin{equation}
\chi(z, \Omega_m, \Omega_{\Lambda}, \Omega_k,
\tilde{\alpha}) =\int_z^{\infty}{\frac{dz}{r^2 g(z)}}= \int_z^{\infty}{\exp{
\left [ -\frac{\frac{d^2z}{d\lambda^2}} {\left ( \frac{dz}{d\lambda} \right
)^2} \right ]} r^{-2} dz}, \label{eq:chi}
\end{equation}
so it is connected with the general solution of Eq.(\ref{eq: stdr2}) and the general expression for  $g(z)$ (see Eq.\ref{eq: scaledr}). Since the metric is conformally stationary, from the Fermat
principle, we get the cosmological lens equation\,:
\begin{equation}
\vec{\beta} = \vec{\theta}-\frac{2R_s}{cH_0}(1 + z_d)[ \chi(z_d) - \chi(z_s)
] \frac{\partial{\psi}}{\partial{\vec{\theta}}} \ ,
\label{eq: lenseq}
\end{equation}
where $R_s$ is the Schwarzschild radius of the deflector.
Denoting \,$\vec{\xi} = D_d \vec{\theta}$, $\vec{\eta} = D_s
\vec{\beta}$, $\vec{\alpha}(\vec{\xi}) =
\displaystyle{\frac{2R_s}{D_d}\frac{\partial{\psi}}{\partial{\vec{\theta}}}}$,
we obtain
\begin{equation}
\vec{\eta}= \frac{D_s}{D_d} \vec{\xi} - D_{ds} \vec{\alpha}(\vec{\xi}) \ ,
\label{eq: lenseq2}
\end{equation}
which is formally identical to the lens equation obtained for $z << 1$.

Let us first discuss the interesting range of the parameter space. From
the available  observations and theoretical considerations it follows that
the radius of curvature of the universe is very large or equivalently that
dimensionless curvature parameter $\Omega_k$ is very small. In our
considerations we assume that $|\Omega_k|\leq 0.05$. Unfortunately no
reliable estimates of $\tilde \alpha$ exist, but in the considered range
of redshifts $z\epsilon (0, 100)$,  we consider the range of
$\tilde\alpha$ $\in$ $[0.3,1] $ (not completely clumpy Universe). We also
allow different values of $\Omega_{\Lambda}$ and with special attention we
treat the case $\Omega_{\Lambda}=0.65$ which seems to be singled out by
observations (\cite{Koch96}, \cite{Koch96b}, \cite{Koch98},
\cite{Perl97},\cite{Tom1}). The density parameters satisfy the constraint
$\Omega_m+\Omega_{\Lambda}+\Omega_k=1$.

In the considered range of redshifts and of the other cosmological
parameters we propose to use the following form of the approximate function
for $r$
\begin{equation}
r(z) =\frac{1}{\sqrt{d_1 + d_2(d_3 + d_4/z + z)^2}},
\label{eq: apprr}
\end{equation}
where the parameters $d_1, d_2, d_3, d_4$ depend on the cosmological
parameters and they are determined by fitting Eq.(\ref{eq: apprr}) to the
corresponding numerical solution. To optimize the fit we use a non linear
regression procedure available in $\it{Mathematica~~ 4.0}$.

\begin{figure}[ht]
\centering
\epsfig{figure=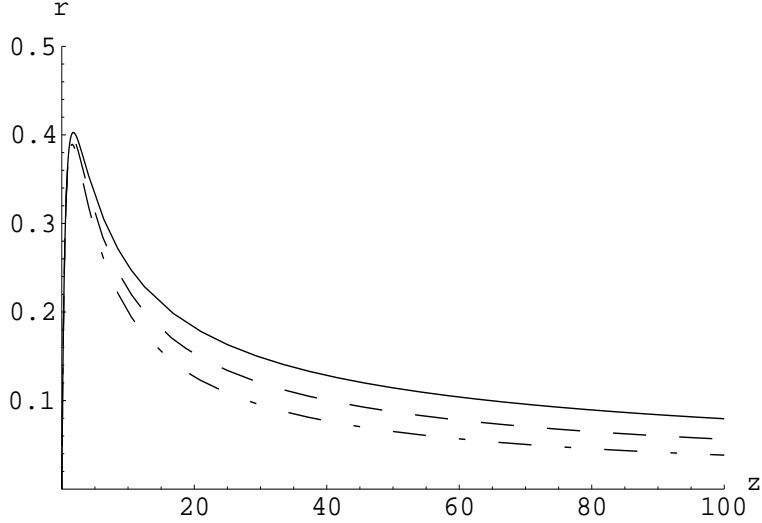,height=7cm,width=0.8\textwidth,clip=}
\caption{$r(z)$ for different values
of $\tilde{\alpha}$, and for fixed $\Omega_{\Lambda }= 0.65$ is shown. The
values of  $\tilde{\alpha}$ are $\tilde{\alpha}=0.9$, $\tilde{\alpha}=0.8$,
$\tilde{\alpha}=0.7$ respectively, $\tilde{\alpha}$ decreases from the
bottom curve up.}
\end{figure}

\begin{figure}[ht]
\centering
\epsfig{figure=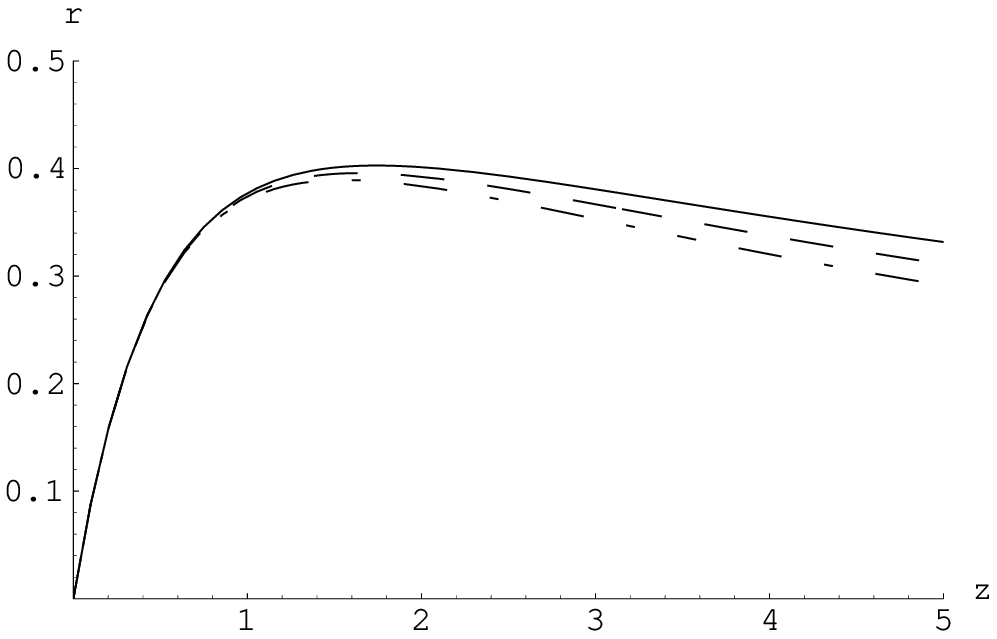,height=7cm,width=0.8\textwidth,clip=}
\caption{Same as in Fig. 3; the z- range is now($0$,$5$)}
\end{figure}

\begin{figure}[ht]
\centering
\epsfig{figure=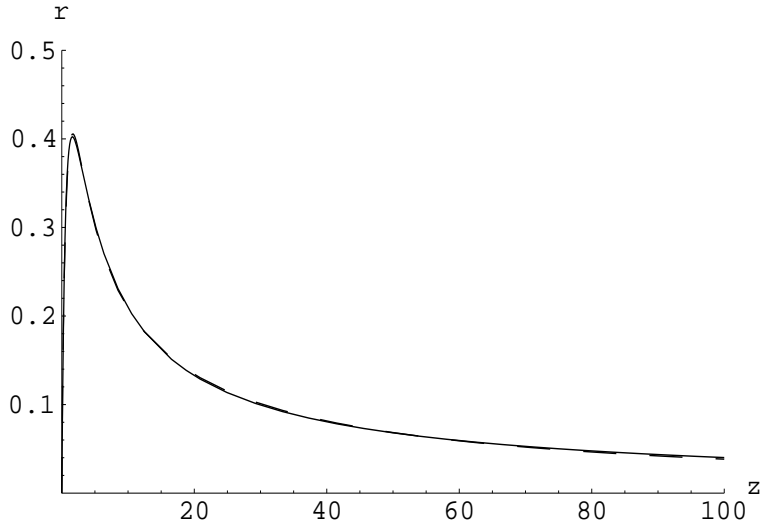,height=7cm,width=0.8\textwidth,clip=}
\caption{The numerical solution of
the Dyer-Roeder equation and our approximated function for fixed
$\Omega_{\Lambda }= 0.7$ and $\tilde{\alpha}=0.9$ are shown. As it is
apparent, the approximating function  works  very well. }
\end{figure}

\begin{figure}[ht]
\centering
\epsfig{figure=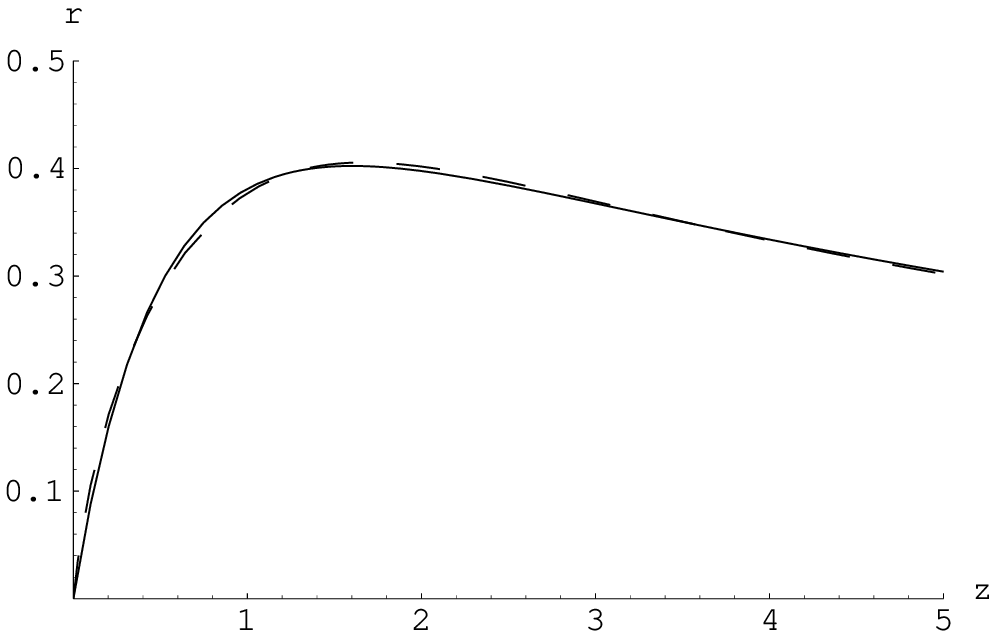,height=7cm,width=0.8\textwidth,clip=}
\caption{Same as in Fig. 5, the z-range is now ($0$, $5$.) }
\end{figure}

\begin{figure}[ht]
\centering
\epsfig{figure=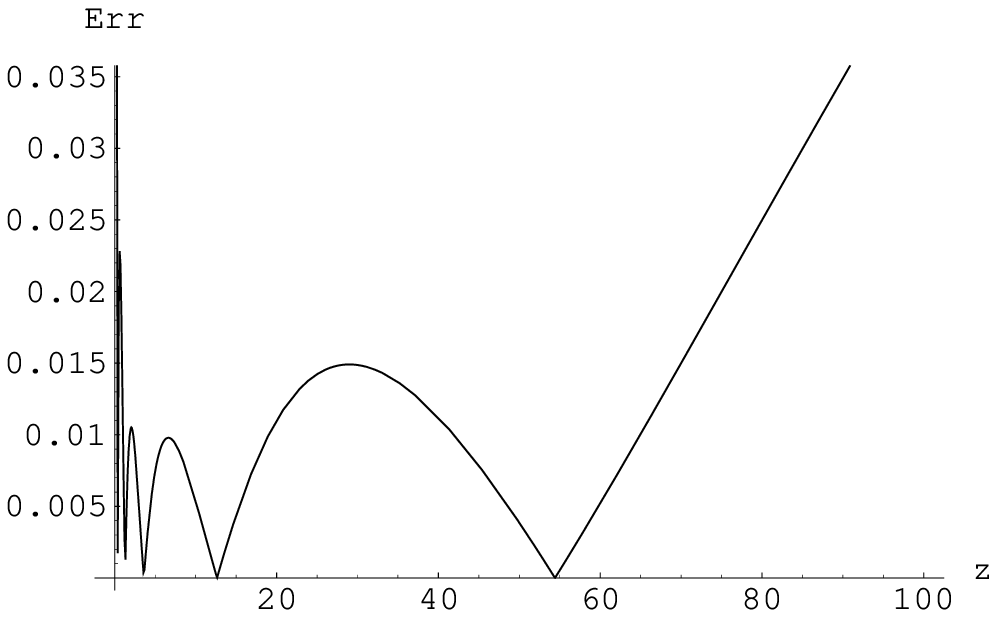,height=7cm,width=0.8\textwidth,clip=}
\caption{The relative error function between the numerical solution of
the Dyer-Roeder equation and our approximated function,  $\Omega_{\Lambda
}= 0.7$, $\tilde{\alpha}=0.9$. }
\end{figure}

We use the same procedure to find the approximate
function $\chi(z)$ which we take in the form
\begin{equation}
\chi(z)={1\over {(e_1+{e_2}z+{e_3}z^2)}},
\label{eq: apprchi}
\end{equation}
where the parameters $e_1, e_2, e_3$ are determined by fitting $\chi(z)$ to
numerically obtained exact values.

\begin{figure}[ht]
\centering
\epsfig{figure=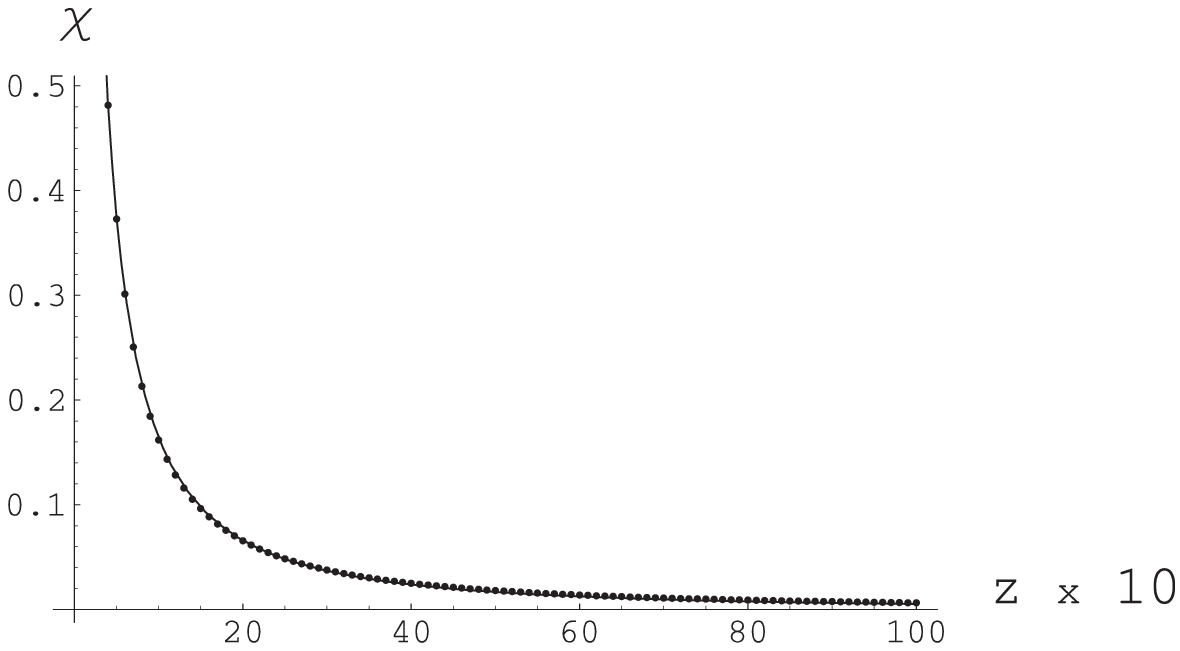,height=7cm,width=0.8\textwidth,clip=}
\caption{This figure shows how our approximate analytical form for
 $\chi$ interpolates
the numerical values. The values of non normalized chi-squared of this
regression is $\chi^2=0.013$. The independent variable is $z{\cdot}10$.}
\end{figure}

We are interested in finding an analytical approximation of the
Dyer-Roeder distances for cosmological values of redshifts (say, for
example, $z\geq0.05)$; in this region the polynomial, which appears in the
denominator of the approximate function $\chi(z)$ does not have zeros. In
Fig.5
- 6 we show the exact angular diameter distance and the fitted approximate
relation for the same value of $\tilde{\alpha}$ and other cosmological
parameters. (The calculations have been done for fixed $\tilde{\alpha}$,
$\Omega_{\Lambda}=0.65$ and $\Omega_k=0$.) In Fig.7 we see that the
approximate angular diameter distance reproduces the exact curve with a
very good accuracy, the error is not larger than $\sim 1\%$, for $z$ in
the range (0, 50) and it is still good for larger $z$ in the range (50,
100). In Fig.8 we show exact numerically obtained function $\chi(z)$ and
the fitted approximate relation for the same value of $\tilde{\alpha}$ and
other cosmological parameters.

Finally we have analyzed how the typical maximum present in the angular
diameter distance depends on the two cosmological parameters
($\Omega_{\Lambda}$, $\tilde{\alpha}$). Actually we perform this analysis
both for the maximum of that distance $r_{max}$ and for the z-value for
which that maximum occurs $z_{max}$. For $\Omega_k=0$ and
$\Omega_{\Lambda}=const.$, $r_{max}$ depends on $\tilde{\alpha}$ as:
\begin{eqnarray}
&&r_{max}(\tilde{\alpha},\Omega_{\Lambda}=const.,\Omega_k=0)=\nonumber\\&&\\
&=&\exp[\zeta_0 +
\zeta_1(\Omega_{\Lambda})+\zeta_2(\Omega_{\Lambda})^2+ \zeta_3(\Omega_{\Lambda})^3]\nonumber
\end{eqnarray}
If we fix $\tilde{\alpha}=const.$ and consider the dependence of $r_{max}$
on $\Omega_{\Lambda}$, we find:
\begin{eqnarray}
&&r_{max} (\tilde{\alpha}=const.,\Omega_{\Lambda},\Omega_k= 0)=
\nonumber\\&&\\
&=&\exp[\varepsilon_0+
\varepsilon_1(\Omega_{\Lambda})+\varepsilon_2(\Omega_{\Lambda})^2+
\varepsilon_3(\Omega_{\Lambda})^3],\nonumber
\end{eqnarray}
where the coefficients $\varepsilon_i$ depend on the clumpiness parameter
$\tilde{\alpha}$ and on $\Omega_k$.  We can also study how $z_{max}$
depends on $\tilde{\alpha}$ and $\Omega_{\Lambda}$. For
$\tilde{\alpha}=const.$, $\Omega_k=0 $, we have that
\begin{eqnarray}
&&z_{max}(\tilde{\alpha}=const.,\Omega_{\Lambda},\Omega_k=0)=\nonumber\\
&&\label{eq:zmax1}\\&=&\exp[\tau_0+\tau_1
(\Omega_{\Lambda})+\tau_2(\Omega_{\Lambda})^2 +
\tau_3(\Omega_{\Lambda})^3+ \tau_4 (\Omega_{\Lambda})^4+ \tau_5(\Omega_{\Lambda})^5].\nonumber
\end{eqnarray}
The coefficients  $\tau_i$ depend on the clumpiness parameter
$\tilde{\alpha}$ and on $\Omega_k$. For $\Omega_{\Lambda}=const.$,
$\Omega_k=0 $, we find
\begin{eqnarray}
 &&z_{max}(\tilde{\alpha},\Omega_{\Lambda}=const.,
\Omega_k=0)=\nonumber\label{eq:zmax2}\\&&\\
&=& \exp[\gamma_0 + \gamma_1 \tilde{\alpha}+ \gamma_2 (\tilde{\alpha})^2+
\gamma_3 (\tilde{\alpha})^3+\gamma_4 (\tilde{\alpha})^4 + \gamma_5 (\tilde{\alpha})^5].
\nonumber
\end{eqnarray}
where the $\gamma_i $ coefficients depend on and the parameters
$\Omega_{\Lambda}$ and $\Omega_k$. From Eq. (\ref{eq:zmax2}) it follows
that for $\Omega_{\Lambda}
= 0.65$ and $\tilde{\alpha}=0.9$, we get $z_{max}=1.62$. When
$\Omega_{\Lambda}=0$, $\Omega_{k}=0$, as it follows from Eq. (\ref{eq:
SEFsol})  $z_{max}=1.25$. It is interesting to note that
$z_{max}(\Omega_{\Lambda}=0, \Omega_{k}=0,
\tilde{\alpha}=0.9)< z_{max}(\Omega_{\Lambda}=0.65,
\Omega_{k}=0,\tilde{\alpha}=0.9)$,
which shows how relevant is the role of the cosmological constant.

\section{Conclusion}

In this paper we have considered the cosmological lens equation in the
case when the cosmological constant $\Lambda\not=0$ and the curvature of
space is different for zero ($k\not=0$). We have included the effects of
non homogeneous distribution of matter which are described by a
phenomenological parameter $\tilde{\alpha}$. Unfortunately at the moment
there are no generally accepted  models that  describe the distribution of
baryonic and dark matter at high redshifts and therefore the influence of
non homogeneities of matter distribution can be included only at this
approximate level.  Following the standard procedure (\cite{SEF}) we use
the Dyer-Roeder distance to find the distance between two objects with
redshifts $z_{1}$ and $z_{2}$.

To find the general solution we slightly enlarged the parameter space by
considering the non zero cosmological constant, so  the cosmological model
is described by the following parameters $\Omega_{\Lambda}\not=0$,
$\Omega_{m}\not=0$, $\tilde{\alpha}\in [0.3,1]$ (not completely clumpy
Universe), $p=0$ (no radiation). The solution that we have found is a
functional combination of the solution given in SEF for the case when
$\Omega_{\Lambda}=0$, $\Omega_{k}=0$ and two hypergeometric functions. Fig.
1 shows that in the flat universe the Dyer-Roeder distance increases for
increasing values of the cosmological constant. In the next step we allow
the curvature of space to be different from zero. This further complicates
the Dyer-Roeder equation which becomes of a Fuchsian type with 4 regular
singular points and one regular singular point at infinity. The general
form of this kind of ordinary differential equations is given in terms of
the P-Riemann symbol.

The obtained exact solution is so complicated that it is practically
useless in practical applications. Therefore we have looked for an
approximate analytic solution simple enough to be used in many
applications and at the same time sufficiently accurate at least in the
interesting range of redshifts. The proposed form of the approximate
solution of the Dyer-Roeder equation depends on four arbitrary parameters.
To fix values of these parameters we fit the approximate solution to the
exact one with the help of a non linear regression method (see {\it
Mathematica~~ 4.0}). Of course, the parameters depend on the density
parameters, but their values for different cosmological parameters can be
tabulated. Following SEF we have also found the function $\chi$ which
appears in the expression for time delays as well as in the lens equation.
We have also proposed approximate analytical form of the function $\chi$
which contains three arbitrary parameters. To find values of these
parameters we use the same method as above.

We have also studied how the redshift corresponding  to the maximal
angular diameter distance depends on the basic parameters determining the
cosmological model.

Let us now consider the following three combinations of the DR
distance:
\begin{eqnarray}
\circ&&\frac{D_{LS}}{D_{OS}}
=\frac{H_0}{c}\frac{r_{LS}}{r_{OS}},\nonumber\\
\circ&&\frac{D_{OL}D_{LS}}{D_{OS}}=\frac{H_0}{c}\frac{r_{OL}r_{LS}}{r_{OS}},\nonumber\\
\circ&&\frac{D_{OL}D_{OS}}{D_{LS}}=\frac{H_0}{c}\frac{r_{OL}r_{OS}}{r_{LS}}.\nonumber
\end{eqnarray}
We have selected these combinations because of the role they play,
respectively, in bending of light, lensing statistics, and time delay.
Asada (\cite{AS97}, \cite{AS98a}, \cite{AS98b}) have found that:
\begin{eqnarray}
\circ&&\frac{D_{LS}}{D_{OS}} (\alpha_1)<\frac{D_{LS}}{D_{OS}}(\alpha_2),\,\,\,
{\rm{for}} \,\,\,\alpha_1< \alpha_2, \nonumber\\
\circ&&\frac{D_{OL}D_{LS}}{D_{OS}}
(\alpha_1)< \frac{D_{OL}D_{LS}}{D_{OS}}(\alpha_2),\,\,\,\,
{\rm{for}}\,\,\,\,
\alpha_1<\alpha_2,\nonumber\\
\circ&&\frac{D_{OL}D_{OS}}{D_{LS}} (\alpha_1)>
\frac{D_{OL}D_{OS}}{D_{LS}}(\alpha_2),\,\,\,\, {\rm{for}}\,\,\,\,
\alpha_1 <\alpha_2. \nonumber
\end{eqnarray}
Using our approximate relations for $r$ and $\chi$ we obtain that:
\begin{eqnarray}
\circ \,\,\,\,\,\ \frac{D_{LS}}{D_{OS}} &=& (1+z_L) r(z_L)
(\chi(z_L)-\chi(z_S))=\nonumber\\
&=&(1+z_L) \frac{1}{\sqrt{d_1 + d_2(d_3 +
d_4/z_L + z_L)^2)}}(\frac{1}{(e_1 + e_2  z_L + e_3
z_L^2)}-\nonumber\\&&\frac{1}{(e_2 + e_2 z_S + e_3z_S^2 )}),\nonumber
\\&&\nonumber\\
\circ\,\,\,\,\,\frac{D_{OL}D_{LS}}{D_{OS}}&=& (1+z_L) r^2(z_L) (\chi(z_L)-\chi(z_S)) =
\nonumber\\&=&(1+z_L) \frac{1}{(d_1 + d_2(d_3 + d_4/z_L + z_L)^2)}(\frac{1}{(e_1 + e_2 z_L
+ e_3 z_L^2)}-\nonumber\\&&\frac{1}{(e_1 + e_2 z_S + e_3 z_S^2)}),\nonumber
\\&&\nonumber\\
\circ\,\,\,\, \frac{D_{OL}D_{OS}}{D_{LS}} &=&\frac{1}{(1+z_L)(\chi(z_L)-\chi(z_S))}
 =\nonumber\\&=&\frac{1}{(1+z_L)}\frac{1}{(\frac{1}{(e_1 + e_2  z_L + e_3 z_L^2)}-
\frac{1}{(e_1 +e_2  z_S + e_3 z^2)})}\nonumber,
\end{eqnarray}
Using the fitted values for the parameters $(d_{1}, d_{2}, d_{3}, d_{4})$
and $(e_{1}, e_{2}, e_{3})$ we confirm results of Asada for
$\tilde{\alpha}$ in the considered range $1\geq {\tilde{\alpha}} \geq
0.3$. \\
Finally we like to stress that from our analysis it follows that
variations in the angular diameter distance caused by the presence of
cosmological constant are quite similar to variations due to changes in
the value of $\tilde{\alpha}$.

\begin{figure}[ht]
\centering
\epsfig{figure=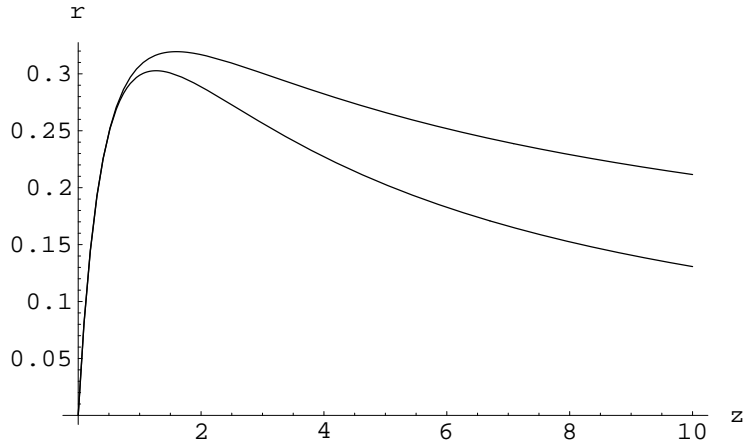,height=6cm,width=0.8\textwidth,clip=}
\caption{r(z) for
$\Omega_{\Lambda}=0$ and $\tilde{\alpha}=0.6$ (upper curve) and
$\Omega_{\Lambda}=0.7$ and $\tilde{\alpha}=1$ (bottom curve). This figure
shows that non homogeneous distribution of matter can mimic the effect of
non zero cosmological constant.}
\end{figure}

 Actually, in Fig.9,  we consider two plots
corresponding to  the values $\Omega_\Lambda=0$ and $\tilde{\alpha}\ne 0$
and $\Omega_\Lambda\ne 0$ and $\tilde{\alpha}=1$ and we see that the
$\tilde{\alpha}$ parameter could mimic the effect of a non zero
cosmological constant. This is an important observation in view of the
recent observational results concerning the non zero value of the
cosmological constant (\cite{Koch96}, \cite{Koch98}, \cite{Koch96b},
\cite{Perl97}). In the future we would like to extend our work and include
also the radiation density parameter.

    ~\\{\it Acknowledgments}.

It is a pleasure to thank V.F. Cardone , G. Covone. C. Rubano and P.
Scudellaro for discussions we had on the manuscript. RdR and AAM are
financially sustained by the M.U.R.S.T. grant PRIN97 ``SIN.TE.SI`` , MD by
the grant 2-P03D-014-17 of the Polish State Committee for Scientific
Research, and EP by the Social European Committee.

{}

\newpage
\section{Appendix A}

\begin{eqnarray}
A& = & -\frac{3\tilde{\alpha}\Omega_m}{2 a_1 a_2 a_3},\nonumber \\
&&\nonumber\\B &
= &
-{3\tilde{\alpha}\Omega_m\over 2\alpha_9}{\times}
 {-\alpha_5(\alpha_3-\alpha_4)+
\alpha_7(\alpha_1-\alpha_4)-\alpha_8{\times}(\alpha_1-\alpha_3)\over
(-\alpha_6(\alpha_3-\alpha_4)+
\alpha_7(\alpha_2-\alpha_4)-\alpha_8(\alpha_2-\alpha_3))},\nonumber \\ &
& \nonumber \\ C & = & -(A+B+D), \\ & & \nonumber \\ D & = &
{(A(\alpha_1-\alpha_3)+B(\alpha_2-\alpha_3))\over
(\alpha_3-\alpha_4)}.\nonumber
\end{eqnarray}

The coefficients $\tilde{A}$,  $\tilde{B}$,  $\tilde{C}$,  $\tilde{D}$ are

\begin{eqnarray}
 && \tilde{A} = -\frac{3}{2}\frac{\Omega_m}{a_1a_2a_3},\nonumber\\
& & \nonumber \\ &&\tilde{B}  =  \frac{-\alpha_3 \alpha_5 \Omega_m+3
\alpha_4 \alpha_5 \Omega_m}{2\alpha_9 (\alpha_3 \alpha_6-\alpha_4\alpha_2\alpha_7+
\alpha_2\alpha_8- \alpha_2\alpha_8)} + \nonumber\\
& &  \nonumber \\ & & + {3\alpha_1
\alpha_7 \Omega_m-3 \alpha_4 \alpha_7 \Omega_m+
3\alpha_1 \alpha_8\Omega_m \over
(2\alpha_9(\alpha_3\alpha_6-\alpha_4\alpha_2\alpha_7+\alpha_2\alpha_8-
\alpha_2\alpha_8))} + \nonumber\\ & &  \nonumber \\ & & +
{2\alpha_7\alpha_9\Omega_k- 2\alpha_8\alpha_9\Omega_k\over(
2\alpha_9(\alpha_3\alpha_6-\alpha_4\alpha_2\alpha_7+\alpha_2\alpha_8-
\alpha_2\alpha_8))}, \\ & &  \nonumber \\ &&\tilde{C} =
\frac{-\alpha_2\alpha_5\Omega_m+3\alpha_4 \alpha_5\Omega_m
}{2\alpha_9
(\alpha_3\alpha_6-\alpha_4\alpha_6-\alpha_2\alpha_7+\alpha_4\alpha_7+
\alpha_2\alpha_8-\alpha_3\alpha_8)}- \nonumber\\
& &  \nonumber \\ & & -
{3\alpha_1\alpha_6\Omega_m+3\alpha_2\alpha_6\Omega_m
\over(2\alpha_9(\alpha_3
\alpha_6-\alpha_4\alpha_6-\alpha_2\alpha_7+\alpha_4\alpha_7+
\alpha_2\alpha_8-\alpha_3\alpha_8))} + \nonumber \\
& &  \nonumber \\ & & +
{3\alpha_1\alpha_8\Omega_m-3\alpha_2\alpha_8\Omega_m-
2\alpha_6\alpha_9\Omega_k+2\alpha_8
\alpha_9\Omega_m\over
(2\alpha_9(\alpha_3\alpha_6-\alpha_4\alpha_6-\alpha_2\alpha_7+\alpha_4\alpha_7+
\alpha_2\alpha_8-
\alpha_3\alpha_8))},\nonumber \\
& &  \nonumber \\ &&\tilde{D} =
-\frac{3\alpha_2\alpha_5\Omega_m-3\alpha_3\alpha_5{\cdot}
\Omega_m+3\alpha_3\alpha_6\Omega_m}{2\alpha_9(-\alpha_3\alpha_6+
\alpha_4\alpha_6+\alpha_2\alpha_7-\alpha_4\alpha_7
-\alpha_2\alpha_8+\alpha_3\alpha_8)} + \nonumber \\
& &  \nonumber \\ & & + { 3\alpha_1\alpha_7\Omega_m-
3\alpha_2\alpha_7\Omega_m-2\alpha_6\alpha_8\Omega_m-2\alpha_6
\alpha_9\Omega_k \over(
2\alpha_9(-\alpha_3\alpha_6+\alpha_4\alpha_6+\alpha_2\alpha_7-\alpha_4\alpha_7
-\alpha_2\alpha_8+\alpha_3\alpha_8))} + \nonumber \\
& &  \nonumber \\ & & + {
2\alpha_8\alpha_9\Omega_k+2\alpha_7\alpha_9\Omega_m \over( 2\alpha_9
(-\alpha_3\alpha_6+\alpha_4\alpha_6+\alpha_2\alpha_7-\alpha_4\alpha_7
-\alpha_2\alpha_8+\alpha_3\alpha_8))}.\nonumber
\end{eqnarray}

The functions $\alpha_i $ being:
\begin{eqnarray}
&&\alpha_1=a_1a_2+a_1a_3+a_2a_3 ,\nonumber\\ &&\alpha_2=a_2a_3, \nonumber
\\ &&\alpha_3=a_1a_3,
\nonumber \\
&&\alpha_4=a_1a_2, \nonumber \\ &&\alpha_5=a_1+a_2+a_3,\nonumber\\
&&\alpha_6=a_2+a_3,\nonumber \\&&\alpha_7=a_1+a_3,
\nonumber\\ &&\alpha_8=a_1+a_2, \nonumber \\ &&\alpha_9=a_1a_2a_3.
\nonumber
\label{eq:costanti}
\end{eqnarray}

\newpage
\section{Appendix B}
We give the form of the coefficient appearing in equations (\ref{eq:
apprr}) and (\ref{eq: apprchi}) , in function of the cosmological
parameter. We show the dependence of the coefficient ${d_i} $ on the
clumpiness parameters $\tilde{\alpha}$, fixing $\Omega_{\Lambda }=0.65 $

\begin{eqnarray}
&&d_1=\exp[a_{d1} + b_{d1}  \tilde{\alpha} + c_{d1}
\tilde{\alpha}^2 + d_{d1} \tilde{\alpha}^3],\nonumber \\ &&\nonumber\\ &&d_2=
\exp{[a_{d2} + b_{d2}   \tilde{\alpha} + c_{d2}
\tilde{\alpha}^2 + d_{d2} \tilde{\alpha}^3]},\nonumber \\ &&\nonumber\\
&& d_3=\exp{[a_{d3} + b_{d3}   \tilde{\alpha} + c_{d3}
\tilde{\alpha}^2 + d_{d3} \tilde{\alpha}^3]},\nonumber\\ &&\nonumber \\&& d_4=
\exp{[a_{d4} + b_{d4}   \tilde{\alpha} + c_{d4}
\tilde{\alpha}^2 + d_{d4} \tilde{\alpha}^3]}.
\nonumber  \label{eq: trenta4}
\end{eqnarray}

Dependence of ${d_i}$ functions from  $\Omega_{\Lambda}$ with
\,$\tilde{\alpha}= 0.8$
\begin{eqnarray}
&&d_1 = \exp{[p_{d1}+ q_{d1} \Omega_{\Lambda} + r_{d1}(\Omega_{\Lambda})^2
+ s_{d1} (\Omega_{\Lambda})^3]},\nonumber \\ &&\nonumber\\
 &&d_2= \exp{(p_{d2}+ q_{d2} \Omega_{\Lambda} + r_{d2}(\Omega_{\Lambda})^2
+ s_{d2} (\Omega_{\Lambda})^3)}, \nonumber\\ &&\nonumber\\
&&d_3=\exp{(p_{d3}+ q_{d3} \Omega_{\Lambda} + r_{d3}(\Omega_{\Lambda})^2 +
s_{d3} (\Omega_{\Lambda})^3)},\nonumber\\ &&\nonumber\\
&&d_4=\exp{[p_{d4}+ q_{d4}
\Omega_{\Lambda} + r_{d4}(\Omega_{\Lambda})^2 + s_{d4}
(\Omega_{\Lambda})^3]}, \nonumber
\label{eq: trenta5}
\end{eqnarray}
\\being  ${\mathsf{P}}_n(\Omega_{\Lambda})$  a polynomial of n-degree in
\,\,$\Omega_{\Lambda}$.\\ We see below the dependence of the
functions ${e_i}$ from $\tilde{\alpha}$ with $\Omega_{\Lambda }= 0.65$
\begin{eqnarray}
&&e_1 = m_{e1} \,+\, n_{e1} \,   \tilde{\alpha} \,+\, o_{e1}\,
\tilde{\alpha}^2
\,+\, p_{e1}
 \tilde{\alpha}^3,
\nonumber \\ &&\nonumber\\
&&e_2 =  m_{e2}\, + \,n_{e2} \,  \tilde{\alpha} \,+\, o_{e2}\,
\tilde{\alpha}^2
\,+ p_{e2}\,
 \tilde{\alpha}^3,\nonumber\\
&&\nonumber\\ &&e_3= m_{e3} \,+ \,n_{e3}\,  \tilde{\alpha}. \nonumber
\label{eq: chicoef}
\end{eqnarray}

Here, we show the dependence of the functions ${e_i}$ from $\Omega_{\Lambda
}$,with $\tilde{\alpha}=0.9$, and $\Omega_k=0$: \\
\begin{eqnarray}
&&e_1 = \tilde{m_{e1}} \,+\, \tilde{n_{e1}} \,  \Omega_{\Lambda} \,+\,
\tilde{o_{e1}}\,
 \Omega_{\Lambda}^2
\,+\,  \tilde{p_{e1}}
\Omega_{\Lambda}^3,
\nonumber \\ &&\nonumber\\
&&e_2 =  \tilde{m_{e2}}\, + \,\tilde{n_{e2}} \,\Omega_{\Lambda} \,+\,
\tilde{o_{e2}}\,
\Omega_{\Lambda}^2
\,+ \tilde{p_{e2}}\,
\Omega_{\Lambda}^3,
\nonumber\\
&&\nonumber\\ &&e_3={\mathsf{P}}_q(\Omega_{\Lambda})/(\exp{[\tilde{m_{e3}}
\Omega_{\Lambda}} + \tilde{n_{e3}]}),\nonumber
\label{eq: chicoef2}
\end{eqnarray}
being ${\mathsf{P}}_q(\Omega_{\Lambda})$ a polynomial of q-degree in
$\Omega_{\Lambda}$ depending on the value of $k$ \\

\end{document}